\documentstyle[12pt]{article}
\begin{document}

\author{M. I. Krivoruchenko \\
{\small {\it Institut f\"ur Theoretische Physik, Universit\"at
T\"ubingen, Auf der Morgenstelle 14}}\\
{\small {\it D-72076 T\"ubingen, Germany}}\\
{\small {\it and}}\\
{\small {\it Institute for Theoretical and Experimental Physics
B.Cheremushkinskaya, 25}}\\
{\small {\it 117259 Moscow, Russia $^{1)}$}}}
\title{Analytical Extension of the Frazer-Fulco Unitarity 
Relations for Isovector Nucleon Form Factors \\
to the Complex t-Plane}
\date{}
\maketitle

\begin{abstract}
The unitarity relations for isovector nucleon form
factors in the two-pion approximation for the intermediate states 
are analytically extended in a model independent way from the real 
half-axis $t>4\mu ^2$ ($\mu $ is the pion
mass) to the whole complex $t$-plane. As a result, the explicit 
integral representations for
the nucleon form factors are constructed in terms of the pion form factor
and imaginary parts of the $t$-channel $p$-wave $\pi N$-scattering 
amplitudes at $t<0$. New
method based on the explicit integral representations 
is developed for numerical evaluation of the two-pion contribution 
to the isovector nucleon form
factors. The results of different methods are compared and their 
efficiency is discussed.

\end{abstract}

\newpage

\section{Introduction}

\setcounter{equation}{0}

$ $
\vspace{-1 cm} 

Recent years new precise experimental data were collected for the 
electromagnetic nucleon form factors at low and intermediate momentum 
transfers at spacelike and timelike regions \cite{exp}. The experimental 
progress stimulates development of the phenomenological models for the 
nucleon form factors.

The unitarity relations for isovector nucleon form factors, discovered by 
Frazer and Fulco \cite{FF}, appeared to be the first 
fruitful attempt to provide field-theoretical framework for 
description of the electromagnetic nucleon form factors at low momentum 
transfers. Being combined 
with the requirements of the analyticity, relativistic 
invariance, and crossing 
symmetry of the $\pi N$-scattering amplitudes, the unitarity relations allow to 
evaluate the two-pion contribution to the absorptive pars of the nucleon 
form factors at the cut $(4\mu ^2,+\infty )$ using experimental 
data on the 
$\pi \pi $-and $\pi N$-scattering amplitudes only. Similar 
relations based on the 
unitarity are used in the analysis of the electromagnetic pion form 
factor \cite{BD}. The unitarity relations explain the 
origin of the vector-meson dominance (VMD) in the pion and nucleon 
form factors. The VMD models are in wide use in the modern 
calculations \cite{HEA} - \cite{HPU}.

The differential approximation scheme for the analytical continuation of 
the $\pi N$-scattering amplitudes from the $s$-channel ($\pi N$ $\rightarrow$ $\pi N$) 
to the $t$-channel ($\pi \pi $ $\rightarrow$  $N\bar N $), suggested by Efremov, 
Meshcheryakov and Shirkov \cite{EMS}, was used by Lendel et al. \cite{LLME} 
to evaluate the nucleon form factors from the unitarity relations (for a review see 
Ref. \cite{SSM}). 	

Hšhler and Pietarinen \cite{HP} performed the alternative 
analysis using the prescription by Frazer and Fulco \cite{FF} for calculation of 
the $t$-channel $\pi N$-scattering amplitudes at $t < 0$ and the discrepancy 
function method to extrapolate the amplitudes to the region $t > 4\mu^2$ 
(for a review see Ref. \cite{HBOOK}).

The agreement of these two methods is surprisingly pure. The estimates 
of the two-pion contribution to the isovector nucleon form factor 
$F_{2v}(t = 0)$ are in reasonable agreement, whereas the estimates of the 
form factor $F_{1v}(t = 0)$ differ by a factor of two. There were no attempts 
to analyze the origin of this discrepancy. This might be a reason for the 
unitarity relations are simply ignored during the last two decades in the 
most calculations of the nucleon form factors.

The apparent disagreement of these two methods indicates presence of 
uncertainties whose origin is not yet understood clearly. The purpose of 
this work is to discuss possible sources of the errors and develop new 
method free from ambiguities inherent in the earlier calculations of isovector nucleon form factors.

The unitarity relations by Frazer and Fulco are derived in the two-pion 
approximation for the intermediate states between the photon and the 
$N\bar N $-pair (see Fig.1). Integral representations for the nucleon form 
factors, equivalent to the FF unitarity \cite{FF}, are constructed in 
Refs. \cite{MIKR,MIKW} using the methods of Refs. \cite{FF}, 
\cite{EMS}-\cite{SSM} and Refs. \cite{FF,HP,HBOOK} and assuming 
the validity of the Frazer-Fulco-Gounaris-Sakurai (FFGS) model 
for the pion form factor \cite{FF,GS}. These representations extend the FF unitarity 
from the cut $(4\mu ^2,+\infty )$ to the complex $t$-plane. For the kaon 
form factor, the representation of such a kind is constructed in 
Ref. \cite{MIKK}. In this paper, we extend the FF unitarity to the complex 
$t$-plane in a model independent way and give numerical estimates for 
the nucleon form factors, based on these representations.
 
The outline of the paper is as follows. In 
the next Sect., we construct analytical representations for the nucleon 
form factors valid in the complex $t$-plane, completely equivalent to 
the FF unitarity. In Sect.3, we develop new method for numerical 
evaluation of isovector nucleon form factors, based on these representations. 
In Sect.4, we compare our results with results of the previous works and 
discuss efficiency of different methods for evaluation of the two-pion 
contribution to the nucleon form factors from the unitarity.

\section{Analytical Extension of the Unitarity Relations}

\setcounter{equation}{0}

$ $
\vspace{-1 cm} 

The isovector nucleon form factors are defined by isovector part of the
electromagnetic nucleon current 
\begin{equation}
\label{II.1}j_\mu (p^{\prime },s^{\prime },p,s)_v=\bar u(p^{\prime
},s^{\prime })[F_{1v}(t)\gamma _\mu +\frac 1{2m}F_{2v}(t)i\sigma _{\mu \nu
}q_\nu ]\tau _3u(p,s) 
\end{equation}
where $t=(p^{\prime }-p)^2$, $m$ the nucleon mass, and $\tau _3$ the third
isotopic Pauli matrix. The form factors are normalized by $F_{1v}(0)=1/2$
and $F_{2v}(0)=(\mu _p-\mu _n-1)/2$ with $\mu _p$ and $\mu _n$ being the
proton and neutron magnetic moments (in n.m.). The proton charge is set
equal to unity.

The unitarity relations for nucleon form factors have the form \cite{FF} 
\begin{equation}
\label{II.2}ImF_{iv}(t)=-\frac{k^3}{\sqrt{t}}F^{*}(t)\Gamma _i(t) 
\end{equation}
where $k=\sqrt{t/4-\mu ^2}$, $\mu $ is the pion mass, and 
$$
\Gamma _1(t)=\frac m{p_{-}^2}[-f_{+}^1+\frac t{4\sqrt{2}m}f_{-}^1], 
$$
\begin{equation}
\label{II.3}\Gamma _2(t)=\frac m{p_{-}^2}[f_{+}^1-\frac m{\sqrt{2}}f_{-}^1] 
\end{equation}
with $p_{-}^2=m^2-t/4$. The $f_{\pm}^1(t)$ are the $t$-channel $p$-wave $\pi
N$-scattering amplitudes. The unitarity relations are shown graphically in
Fig. 1.

In the approximation of neglecting all but the two-pion intermediate state,
the values $\Gamma _i(t)$ obey the dispersion relations \cite{FF} 
\begin{equation}
\label{II.4}\Gamma _i(t)/F_\pi (t)=\frac 1\pi \int_{-\infty }^a\frac{%
Im\Gamma _i(t^{\prime })}{t^{\prime }-t}\frac{dt^{\prime }}{F_\pi (t^{\prime
})} 
\end{equation}
with 
\begin{equation}
\label{II.5}a=4\mu ^2-\frac{\mu ^4}{m^2}. 
\end{equation}

Eqs.(2.2) and (2.4) are exactly valid below the four-pion threshold at the
interval $4\mu ^2<t<16\mu ^2$. The branching ratio \cite{PDG} $B(\rho$
$\rightarrow$ $4\pi )<0.002$ is small, so the two-pion approximation for the
unitarity relations is valid with high accuracy at the $\rho $-meson peak
and even at higher energies.

It is useful to define a function 
\begin{equation}
\label{II.6}K_\pi (t)=\frac 1\pi \int_{4\mu ^2}^{+\infty }\frac{t_1}{%
t_1+t^{\prime }}\frac{(t^{\prime }/4-\mu ^2)^{3/2}}{\sqrt{t^{\prime }}}%
|F_\pi (t^{\prime })|^2\frac{dt^{\prime }}{t^{\prime }-t}. 
\end{equation}
with $t_1>0$. This function is analytical in the complex $t$-plane with a
cut $(4\mu ^2,+\infty ).$ The factor $t_1/(t_1+t^{\prime })$ is introduced
to ensure better convergence of the integral. With the help of the function
(2.6) the unitarity relations (2.2) take the form 
\begin{equation}
\label{II.7}ImF_{iv}(t)=-\frac{t_1+t}{t_1}ImK_\pi (t)\Gamma _i(t)/F_\pi (t). 
\end{equation}
Using Eq.(2.4), the once subtracted dispersion integral for the nucleon form
factors can easily be evaluated to give 
\begin{equation}
\label{II.8}F_{iv}(t)=F_{iv}(-t_1)+\frac{t_1+t}{t_1}\frac 1\pi \int_{-\infty
}^aIm\Gamma _i(t^{\prime })\frac{K_\pi (t)-K_\pi (t^{\prime })}{t-t^{\prime }%
}\frac{dt^{\prime }}{F_\pi (t^{\prime })}. 
\end{equation}
For computation of the nucleon form factors, it is sufficient to know the
imaginary parts $Im\Gamma _i(t)$ at $t<a$. The integral in Eq.(2.8),
however, converges very slowly. It is determined by low- and high-energy
parts of the values $Im\Gamma _i(t)$. The high-energy part $-\infty
<t<b$ is unknown. It has been shown \cite{PIET} that there are no objections 
against the truncated partial wave 
expansion in the region - $40\mu ^2 < t < 0$, so we set $b = - 40\mu ^2$. 
The values $\Gamma _i(t)$ can be decomposed to
the Born and rescattering parts: $\Gamma _i(t)=\Gamma _{iB}(t)+\Gamma
_{iR}(t)$. The real and imaginary parts of the values $\Gamma _{iR}(t)$ are
know on the interval $(b,0)$ from the phase-shift analysis of the $\pi N$%
-scattering amplitudes. The problem consists in an approximate evaluation of
the high-energy part with the help of the information on the low-energy real
pars $Re\Gamma _i(t)$.

The once subtracted dispersion representation for the values $\Gamma _i(t)$
has the form 
\begin{equation}
\label{II.9}\Gamma _i(t)/F_\pi (t)=Re\frac{\Gamma _i(\kappa )}{F_\pi (\kappa
)}+\frac{t-\kappa }\pi P\int_{-\infty }^a\frac{Im\Gamma _i(t^{\prime })}{%
(t^{\prime }-\kappa )(t^{\prime }-t)}\frac{dt^{\prime }}{F_\pi (t^{\prime })}%
. 
\end{equation}
The value $\kappa \ $is assumed to be real, $\kappa <$ $a$. The analog of
Eq.(2.8) reads 
$$
F_{iv}(t)=F_{iv}(-t_1)+\frac{t_1+t}{t_1} 
$$
\begin{equation}
\label{II.10}(Re\frac{\Gamma _i(\kappa )}{F_\pi (\kappa )}+\frac 1\pi
P\int_{-\infty }^a\frac{Im\Gamma _i(t^{\prime })}{(t^{\prime }-\kappa )}%
\frac{(t-\kappa )K_\pi (t)-(t^{\prime }-\kappa )K_\pi (t^{\prime })}{%
(t-t^{\prime })}\frac{dt^{\prime }}{F_\pi (t^{\prime })}). 
\end{equation}
In comparison with Eq.(2.8), the integral at high $t$ is suppressed by the
additional power of $t$. Eqs.(2.8) and (2.10) constitute the main result of
this section.

These equations can be simplified when the FFGS model$\ $\cite{FF,GS} for
the pion form factor is applied. In this model, the pion form factor has the
form 
\begin{equation}
\label{II.11}F_\pi (t)=\frac{t_0+t}{t_0}\frac{D(0)}{D(t)} 
\end{equation}
where 
\begin{equation}
\label{II.12}D(t)=A+Bt+k^2h(t) 
\end{equation}
is the $D$-function. It determines the phase shift $\delta _{11}(t)$ in the $%
I=J=1$ $\pi \pi $-partial wave. The exponentially large parameter $t_0\propto \mu
^2\exp (\frac{8\pi k_\rho ^3}{m_\rho ^2\Gamma _\rho })$ is a unique root of
equation $D(-t_0)=0$. Here $m_\rho $ and $\Gamma _\rho $ are the $\rho $%
-meson mass and width, $k_\rho =\sqrt{m_\rho ^2/4-\mu ^2}$. The parameters $%
A $ and $B$ are fixed by requirements%
$$
\begin{array}{c}
\delta _{11}(m_\rho ^2)=\pi /2, \\ 
\delta _{11}^{\prime }(m_\rho ^2)=1/(m_\rho \Gamma _\rho ). 
\end{array}
$$
In this way one gets 
$$
A=m_\rho ^2+\mu ^2h_\rho +m_\rho ^2k_\rho ^2h_\rho ^{\prime }, 
$$
\begin{equation}
\label{II.13}B=-1-\frac 14h_\rho -k_\rho ^2h_\rho ^{\prime }. 
\end{equation}

The function $h(t)$ is determined at $t<0$ by expression 
\begin{equation}
\label{II.14}\frac{k_\rho ^3}{m_\rho ^2\Gamma _\rho }h(t)=\frac 1{2\pi }%
\sqrt{\frac{4\mu ^2-t}{-t}}\ln (\frac{\sqrt{4\mu ^2-t}+\sqrt{-t}}{\sqrt{4\mu
^2-t}-\sqrt{-t}}). 
\end{equation}
The analytical continuation of this function to the complex $t$-plane has a
cut at $t>4\mu ^2$. At the interval $0<t<4\mu ^2$ the function $h(t)$ is
analytical and real. At $t>4\mu ^2$%
$$
Reh(t)=\frac{t-4\mu ^2}th(t-4\mu ^2), 
$$
\begin{equation}
\label{II.15}\frac{k_\rho ^3}{m_\rho ^2\Gamma _\rho }Imh(t+i0)=-\frac k{%
\sqrt{t}}. 
\end{equation}
In Eqs.(2.7) $h_\rho =Reh(m_\rho ^2)$ and $h_\rho ^{\prime
}=Reh^{\prime }(m_\rho ^2)$. The $D$-function at the origin takes the form 
\begin{equation}
\label{II.16}D(0)=m_\rho ^2+\mu ^2h_\rho +m_\rho ^2k_\rho ^2h_\rho ^{\prime
}-\frac{\mu ^2m_\rho ^2\Gamma _\rho }{\pi k_\rho ^3}. 
\end{equation}
In the narrow width limit, $A=m_\rho ^2$, $B=-1$, $D(0)=m_\rho ^2$, $%
t_0=\infty $, and so the $F_\pi (t)$ takes a monopole form.

The following relation takes place above the two-pion threshold 
\begin{equation}
\label{II.17}\frac{k^3}{\sqrt{t}}|F_\pi (t)|^2=D(0)\frac{k_\rho ^3}{m_\rho
^2\Gamma _\rho }\frac{t_0+t}{t_0}ImF_\pi (t+i0). 
\end{equation}
The dispersion integral (2.6) for $t_1=t_0$ can then be evaluated to give 
\begin{equation}
\label{II.18}K_\pi (t)=D(0)\frac{k_\rho ^3}{m_\rho ^2\Gamma _\rho }F_\pi
(t). 
\end{equation}
Eq.(2.10) becomes \cite{MIKW} 
$$
F_{iv}(t)=F_{iv}(-t_0)-D(0)\frac{k_\rho ^3}{m_\rho ^2\Gamma _\rho }\frac{%
t_0+t}{t_0}F_\pi (t) 
$$
\begin{equation}
\label{II.19}(Re\frac{\Gamma _i(\kappa )}{F_\pi (\kappa )}+\frac 1\pi
P\int_{-\infty }^a\frac{Im\Gamma _i(t^{\prime })}{(t^{\prime }-\kappa )}%
\frac{(t-\kappa )/F_\pi (t^{\prime })-(t^{\prime }-\kappa )/F_\pi (t)}{%
(t^{\prime }-t)}dt^{\prime }). 
\end{equation}

The imaginary part of the ratio $F_{iv}(t)/F_\pi (t)$ has a simple
representation: 
\begin{equation}
\label{II.20}Im\{(F_{iv}(t)-F_{iv}(-t_0))/F_\pi (t)\}/\frac{k^3}{\sqrt{t}}%
=-\frac 1\pi \int_{-\infty }^a\frac{Im\Gamma _i(t^{\prime })}{t^{\prime }-t}%
dt^{\prime }. 
\end{equation}

\section{Numerical Evaluation of the Nucleon Form Factors}

\setcounter{equation}{0}

$ $
\vspace{-1 cm} 

In this Sect., we give numerical estimates for the nucleon form factors
using the FFGS model for the pion form factor.

In the real part of the ratio $F_{iv}(t)/F_\pi (t)$, convergence of the
integral can be improved by taking a weighted sum%
$$
Re\{(F_{iv}(t)-F_{iv}(-t_0))/F_\pi (t)\}=-D(0)\frac{k_\rho ^3}{m_\rho
^2\Gamma _\rho }\frac{t_0+t}{t_0} 
$$
\begin{equation}
\label{(3.1)}\sum_{j=1}^Nc_{ij}(Re\frac{\Gamma _i(\kappa _j)}{F_\pi (\kappa
_j)}+\frac 1\pi P\int_{-\infty }^a\frac{Im\Gamma _i(t^{\prime })}{(t^{\prime
}-\kappa _j)}\frac{(t-\kappa _j)/F_\pi (t^{\prime })-(t^{\prime }-\kappa
_j)Re\{1/F_\pi (t)\}}{(t^{\prime }-t)}dt^{\prime }). 
\end{equation}
with the coefficients 
\begin{equation}
\label{(3.2)}\sum_{j=1}^Nc_{ij}=1. 
\end{equation}

The subtraction points $\kappa _j$ are assumed to belong to the interval $%
b<\kappa _j<0$, so the values $Re\frac{\Gamma _i(\kappa _j)}{F_\pi (\kappa
_j)}$ entering Eq.(3.1) are known. The coefficients $c_{ij}$ can be chosen
such as to suppress the high-energy part $-\infty <t<b$ of the integral. 

The
integral in the right hand side of Eq.(3.1) can be rewritten in the form 
\begin{equation}
\label{(3.3)}\sum_{j=1}^Nc_{ij}Re\frac{\Gamma _i(\kappa _j)}{F_\pi (\kappa
_j)}+\frac 1\pi (\int_{-\infty }^b+P\int_b^a)\sum_{j=1}^Nc_{ij}Im\Gamma
_i(t^{\prime })\phi (t,t^{\prime },\kappa _j)dt^{\prime } 
\end{equation}
where

\begin{equation}
\label{(3.4)}\phi (t,t^{\prime },\kappa _j)=\frac 1{t^{\prime }-\kappa _j} 
\frac{(t-\kappa _j)/F_\pi (t^{\prime })-(t^{\prime }-\kappa _j)Re\{1/F_\pi
(t)\}}{(t^{\prime }-t)} 
\end{equation}
The first and third terms are known. We treat square of the second term as
an error to be minimized. It can be evaluated as%
$$
(\frac 1\pi \int_{-\infty }^b\sum_{j=1}^Nc_{ij}Im\Gamma _i(t^{\prime })\phi
(t,t^{\prime },\kappa _j)dt^{\prime })^2\approx (\frac 1\pi \int_{-\infty
}^b\sum_{j=1}^Nc_{ij}Im\Gamma _{iB}(t^{\prime })\phi (t,t^{\prime },\kappa
_j)dt^{\prime })^2\leq 
$$
\begin{equation}
\label{(3.5)}\frac 1{\pi ^2}\int_{-\infty }^b(Im\Gamma _{iB}(t^{\prime
}))^2dt^{\prime }\int_{-\infty }^b(\sum_{j=1}^Nc_{ij}\phi (t,t^{\prime
},\kappa _j))^2dt^{\prime }=\epsilon _1^2/(D(0)\frac{k_\rho ^3}{m_\rho
^2\Gamma _\rho })^2 
\end{equation}

The second source of the error is finite precision of the rescattering part
of the values $Re\Gamma _{iR}(t)$. The corresponding contribution to the
error of the expression (3.3) reads 
\begin{equation}
\label{(3.6)}\epsilon _2^2=(D(0)\frac{k_\rho ^3}{m_\rho ^2\Gamma _\rho }%
)^2\sum_{j=1}^Nc_{ij}^2(Re\frac{\Gamma _{iR}(\kappa _j)}{F_\pi (\kappa _j)}%
)^2\Delta _j^2 
\end{equation}
where $\Delta _j=0.05$ are relative errors for the $Re\Gamma _{iR}(t)$. When $%
N $ goes to infinity, the values $\Delta _j$ are no longer statistically
independent, in which case expression (3.6) overestimates the error. In this
work, we restrict ourselves by relatively small values $N\leq 12$. The
errors of similar nature coming from the finite precision of the $Im\Gamma
_{iR}(t)$ are neglected.

The conventional minimum of the total error $\epsilon _{tot}^2=\epsilon
_1^2+\epsilon _2^2$ with respect to the coefficients $c_{ij}$ under the
constraint (3.2) imposed can be found by introducing the Lagrange multiplier 
$\lambda $. The coefficients $c_{ij}$ and $\lambda $ are solutions of a
system of the $N+1$ linear equations obtained by taking derivatives with
respect to the $c_{ij}$ and $\lambda $ of the function 
\begin{equation}
\label{III.7}\epsilon _1^2+\epsilon _2^2-\lambda (\sum_{j=1}^Nc_{ij}-1). 
\end{equation}
The coefficients $c_{ij}$ and $\lambda $ are $t$-dependent.

The real and imaginary parts of the rescattering amplitudes $\Gamma _{iR}$
on the interval $(b,0)$ are calculated in Ref. \cite{HP} and more recent
results can be found in Ref. \cite{HBOOK}.

We test first convergence of the method with increasing the number $N$ of
the weight coefficients $c_{ij}$. In Figs. 2(a,b) the values of the form
factors at zero momentum transfer are shown, together with the errors 
$\epsilon _{tot}$, as functions of the number $N$. The subtractions are made
at the points $\kappa _j=b(j-1/2)/N$. For $N>6$ the errors decrease slowly.
The predictions are more stable for the $F_{2v}(0)$. It is seen that
variations of the $F_{1v}(0)$ are within the error bars. 
The variations of the $F_{2v}(0)$ are smaller than
the theoretical errors. The coefficients $c_{ij}$ have
irregular behavior with increasing the $N$. The results shown in Figs.
2(a,b), however, agree with each other for different values of the $N$. This
consistency check is equivalent to the consistency check for the analytical
continuation of the amplitudes $\Gamma _i$ to the region $t>4\mu ^2$ with
the use of the discrepancy method.

The convergence of the integral in Eq.(2.20) for imaginary part of the ratio 
$F_{iv}(t)/F_\pi (t)$ can be improved by adding to the right hand side a
weighted sum of expressions 
\begin{equation}
\label{(3.8)}Re\frac{\Gamma _i(\kappa _j)}{F_\pi (\kappa _j)}-\frac 1\pi
P\int_{-\infty }^a\frac{Im\Gamma _i(t^{\prime })}{t^{\prime }-\kappa _j}%
\frac{dt^{\prime }}{F_\pi (t^{\prime })}=0. 
\end{equation}
The integrals in Eqs.(2.20) and (3.8) have similar structure. We chose the
weights $d_{ij}$ to suppress the high-energy part of the integral (2.20).

The right hand side of Eq.(2.20) takes the form 
\begin{equation}
\label{(3.9)}\sum_{j=1}^Nd_{ij}Re\frac{\Gamma _i(\kappa _j)}{F_\pi (\kappa
_j)}-\frac 1\pi (\int_{-\infty }^b+P\int_b^a)Im\Gamma _i(t^{\prime
})(\sum_{j=1}^Nd_{ij}\frac 1{t^{\prime }-\kappa _j}\frac 1{F_\pi (t^{\prime
})}+\frac 1{t^{\prime }-t})dt^{\prime } 
\end{equation}
To ensure faster convergence of the integral, we require 
\begin{equation}
\label{(3.10)}\sum_{j=1}^Nd_{ij}=0. 
\end{equation}
The integral can be regularized afterwards by a substitution $1/(t^{\prime
}-\kappa _j)$ $\rightarrow$ $1/(t^{\prime }-\kappa _j)-1/(t^{\prime }-\kappa _0)$.

The error coming from the unknown high-energy part of the integral can be
evaluated to be 
\begin{equation}
\label{(3.11)}\delta _1^2=\frac 1{\pi ^2}\int_{-\infty }^b(Im\Gamma
_{iB}(t^{\prime }))^2dt^{\prime }\int_{-\infty }^b(\sum_{j=1}^Nd_{ij}\frac
1{(t^{\prime }-\kappa _j)(t^{\prime }-\kappa _0)}\frac 1{F_\pi (t^{\prime
})}+\frac 1{t^{\prime }-t})^2dt^{\prime } 
\end{equation}

The error coming from the finite precision of the rescattering parts $%
Re\Gamma _{iR}(t)$ reads 
\begin{equation}
\label{(3.12)}\delta _2^2=\sum_{j=1}^Nd_{ij}^2(Re\frac{\Gamma _{iR}(\kappa
_j)}{F_\pi (\kappa _j)})^2\Delta _j^2 
\end{equation}

The conventional minimum of the total error 
$\delta _{tot}^2=\delta_1^2+\delta _2^2$ can be found with the help of 
the Lagrange multiplier 
$\gamma $ corresponding to the constraint (3.10). The coefficients $d_{ij}$
and $\gamma $ are solutions of a system of the $N+1$ linear equations
obtained by setting equal to zero derivatives with respect to the $d_{ij}$
and $\gamma $ of the function 
\begin{equation}
\label{(3.13)}\delta _1^2+\delta _2^2-\gamma \sum_{j=1}^Nd_{ij}. 
\end{equation}

In Figs. 2(c,d) the values $Im\{(F_{iv}(t)\ -\ F_{iv}(-t_0))/F_\pi (t)\}$ at
the $\rho $-meson peak $t=m_\rho ^2$, together with the errors, are shown as
functions of the number $N$. The results seem to converge with increasing 
the $N$. 
The variations are within the error bars. The
predictions are more stable for the $F_{2v}(m_\rho ^2)$ than for the $%
F_{1v}(m_\rho ^2)$. The variations of the $F_{2v}(m_\rho ^2)$ are noticeably
smaller as compared to the errors.

The theoretical errors are defined by replacing $\Gamma_i(t)$
 $\rightarrow$ $\Gamma_{iB}(t)$
in the integral over the region $(-\infty,b)$. These estimates are valid up to 
unknown coefficients. These coefficients can apparently be 
fixed by requiring that the theoretical errors be equal to the statistical 
ones. In case of the form factor $F_{1v}(t)$, such a coefficient is close to 
unity, whereas in case of the form factor $F_{2v}(t)$ such a coefficient 
should be 5-7 times smaller than the unity.

Shown in Figs. 3(a-d) are real and imaginary parts of the ratios between the
nucleon form factors and the pion form factor versus $t$. The errors $%
\epsilon _{tot}$ for the real parts and $\delta _{tot}$ for the imaginary
parts are shown. The visible $t$-dependence of the errors $\delta _{tot}$ in
Figs. 3(c,d) is mainly due to the logarithmic scale of the plots.

\section{Discussions}

\setcounter{equation}{0}

$ $
\vspace{-1 cm} 

In the differential approximation scheme by Efremov, Meshcheryakov and Shirkov
\cite{EMS}, the $t$-channel $p$-wave projections of the $\pi N$-scattering 
amplitudes are calculated with an approximation of the $p$-wave amplitudes 
by linear superpositions of the forward and backward amplitudes and their 
higher derivatives. The analytical continuation to the unphysical region 
$t > 4\mu ^2$ is performed with the help of dispersion relations for the forward 
and backward amplitudes. On the basis of the unitarity relations by Frazer 
and Fulco \cite{FF}, the nucleon form factors are analyzed to the lowest order of 
the differential approximation by Lendel et al. \cite{LLME} (see also Ref. 
\cite{SSM}). 

In the original work by Frazer and Fulco, the following program was suggested 
for evaluation of the nucleon form factors from the unitarity relations: The 
invariant $s$-channel $\pi N$-scattering amplitudes at fixed $t < 0$ are 
continued analytically from the physical region to the unphysical region between 
the $s$- and $u$-channels. Afterwards, the $t$-channel partial wave projections 
of these amplitudes can be computed. The complex $p$-waves $f_{\pm }^1(t)$ are known 
in the region - $26\mu ^2 < t < 0$ in which the partial wave expansion converges. 
It has been shown \cite{PIET} that there are no objections against the truncated 
expansion also in the region - $40\mu ^2 < t < 0$. The analytical continuation 
of the complex $p$-waves $f_{\pm }^1(t)$ to the region $t > 4\mu ^2$ can be 
made with 
the use of the extended unitarity relation, Eq.(2.4), according to which the 
ratios $f_{\pm }^1(t)/F_\pi (t)$ (equivalently $\Gamma _i(t)/F_\pi (t)$) are 
analytical functions in the complex $t$-plane with the left branch cut 
$(-\infty ,a)$. In practice, the discrepancy function method is suitable in a 
restricted region of the complex $t$-plane for solving this task, since the 
high-energy behavior of the scattering amplitudes is unknown. Given that the 
analytical continuation to the region $t > 4\mu ^2$ is performed, one can 
compute the dispersion integral for the nucleon form factors. As $t$ increases, 
the analytical continuation becomes less precise, so the upper limit 
 $\Lambda^2\approx 1$ $GeV^2$ should be introduced to exclude the high-energy 
part of the dispersion integral. This program was performed, within the 
framework of the FFGS model, by Hoehler and Pietarinen \cite{HP} for the 
detailed analysis of the isovector nucleon form factors (see also 
Ref. \cite{HBOOK}).

In the latter approach, one should compute three integrals numerically: The first 
one comes from the computation of the $t$-channel partial wave projections 
of the $\pi N$ amplitudes at $t < 0$. The second one comes from the dispersion 
integral for analytical continuation of the $p$-wave amplitudes $f_{\pm}^1(t)$ 
to the region $t>4\mu ^2$. The third one comes from the dispersion integral 
for the nucleon form factors. 

We demonstrated that the analytical evaluation of the last integral is also 
possible. In doing so, we arrived to the explicit one-dimensional integral 
representation (2.19) for the nucleon form factors in terms of the pion 
form factor and imaginary parts of the amplitudes $\Gamma _i(t)$ at $t < a$. 
There is no therefore need to perform analytical continuation of the amplitudes
$\Gamma _i(t)$ to the region $t > 4\mu ^2$. The representation (2.19) is 
constructed for the FFGS model. The model independent integral 
representations are given by Eqs.(2.8) and (2.10).

In Eq.(2.8), the imaginary part of the amplitudes $\Gamma _i(t)$ occurs
only. Due to the extended unitarity (2.4), the low-energy real parts $%
Re\Gamma _i(t)$ are expressible through the high-energy imaginary parts $%
Im\Gamma _i(t)$. Therefore, the real parts contain information about
high-energy behavior of the amplitudes. This information can be used for
evaluation of the high-energy part of the integrals. In Sect.3, we described
the corresponding numerical algorithm.
	
To get an idea about accuracy of the method \cite{EMS} - \cite{SSM}, one needs 
to calculate the next order term in the differential expansion. This work is 
not done yet. It is clear, 
however, that the Born amplitude should be treated beyond the differential 
approximation method, since the logarithmic singularity at $t = a$ provides 
pure convergence of the series expansion.
	
The errors of the analysis \cite{HP} plotted on Fig.1 are ours. They are 
calculated using the claimed accuracy $\pm 15\%$ in Ref. \cite{HP} for the 
amplitudes $f_{\pm}^1(t)$ at the $\rho $-meson peak. There exist additional 
errors not plotted on Fig.2.
	
(i) The existence of the cut-off parameter $\Lambda ^2\approx 1\;GeV^2$ 
implies that a contribution of the $\rho $-meson tail $t>\Lambda ^2$ to 
the nucleon form factors is disregarded. The results \cite{HP} are $\Lambda$
-dependent. In our approach, the cut-off parameter is safely sent to infinity 
(the integrals converge).
	
(ii) In the analysis \cite{HP}, the contribution from the interval 
-$120\mu ^2 < t < b$, in which the truncated partial wave expansion of the 
amplitudes diverges, is included in the dispersion integral. The partial wave 
expansion gives at the interval -$120\mu ^2 < t < b$ substantial part of the 
discontinuity $Im{\Gamma _i(t)}$. The corresponding errors cannot be controlled, 
 since the values $Im{\Gamma _{iR}(t)}$ calculated from the partial wave 
expansion exhibit at $t < b$ strong tendency to increase with decreasing $t$.

The contribution from the interval $(-\infty ,-120\mu ^2)$ in Ref. \cite{HP} is
interpolated by a polynomial $a+bt$. From the unitarity it follows that 
$\Gamma _1(t)=O(1/t)$ and $\Gamma _2(t)=O(1/t^{3/2})$ as $t\mapsto \infty $. 
Taking into account that in the FFGS model $F _{\pi}(t)=O(1/log(t))$, 
we conclude that
the dispersion integral in Eq.(2.4) can be evaluated, respectively, as $%
O(log^2(t)/t)$ and $O(1/t)$ as $t\mapsto \infty $. The contribution from
the interval $(-\infty ,-120\mu ^2)$ therefore vanishes when $t$ increases.
The polynomial representation $a+bt$ does not satisfy this requirement. The
simple interpolations $a/(b+t)$ and $log((a+t)/(b+t))$ could provide
the correct asymptotic form.

The distinction between the methods \cite{EMS} - \cite{HBOOK} and ours makes 
comparison of the results not so straightforward.

The comparison of the results makes sense for zero 
value of the subtraction 
constant $F_{iv}(-t_0)=0$. The quark counting rules imply that the total 
nucleon form factors, containing all multipion and $N\bar N$ contributions 
to the nucleon 
spectral functions, vanish as $t$ goes to infinity. In the 
two-pion approximation, 
the value $F_{iv}(-t_0)$ should not, however, be strictly equal to zero.

In Fig. 2(a-d), we show our results at two points $t=0$ and $t=m_\rho ^2$ 
for different numbers of the weight coefficients. The errors are theoretical 
ones, $\epsilon _{tot}$ and $(k^3/\sqrt{t})\delta _{tot}$. The results of 
Ref. \cite{HP} are also displayed.
	
The most typical results are summarized in Table 1. Ref. \cite{HP} gives 
for the amplitudes $f_{\pm}^1(t)$ at the $\rho$ -meson peak an error of 
$\pm 15\%$, 
resulting in turn in a small error of the form factor $F_{1v}(0)$, which 
is 4-5 times smaller as compared to our predictions. As we already discussed,
some of the uncertainties in the analysis of Ref. \cite{HP} can hardly be 
controlled, so the errors of Ref. \cite{HP} are greater as compared to 
those shown on Figs. 2 and 3 and in Table 1. Our method for large N gives 
systematically lower (negative) values for the difference 
$F_{1v}(0) - F_{1v}(-t_0)$. The error $\epsilon _{tot}$ is, however, large. 
The unitarity 
relations, therefore, do not allow to give accurate predictions for the form 
factor $F_{1v}(t)$. The value $F_{2v}(0)$ in our approach is well defined. 
The estimates of Refs. \cite{LLME} and \cite{HP} are in reasonable agreement 
with ours. The accuracy of our predictions for the $F_{2v}(t)$ is
better as compared to the earlier ones.

\section{Concluding Remarks}

\setcounter{equation}{0}

$ $
\vspace{-1 cm} 

The unitarity relations by Frazer and Fulco are written in 
the two-pion approximation for imaginary parts of the isovector 
nucleon form factors at the branch cut $(4\mu ^2,+\infty )$. We showed 
that the unitarity relations for the nucleon form factors can be 
solved to give analytical one-dimensional integral representations 
for the nucleon form factors valid in the whole complex $t$-plane 
(Eqs.(2.13) and (2.15)). The nucleon form factors are expressed in 
terms of the pion form factor and the $t$-channel $p$-wave amplitudes 
$\Gamma _i(t)$ at $t < a$.

The representation (2.15) is used to develop new method for numerical 
evaluation of the nucleon form factors from the unitarity relations. 
The method is described in Sect.3. It has some advantages as compared 
to the earlier methods: (i) there is no need to perform analytical 
continuation of the amplitudes $\Gamma _i(t)$ from the region $t < 0$ 
to the region $t > 4\mu ^2$, (ii) no cut off parameter $\Lambda$ to 
be introduced in the dispersion integrals for the nucleon form factors.

The numerical estimates of the nucleon form factors from the unitarity 
relations in the two-pion approximation are at present in the rough 
qualitative agreement for different methods. These estimates are 
considered as efficient and reliable if results of the different 
methods will be brought into the correspondence by evaluation of the 
high-order approximations and careful analysis of possible sources 
of errors.

\vspace{0.5 cm}

{\bf Acknowledgments}

I am grateful to Amand Faessler and A. J. Buchmann for 
discussions of the unitarity relations and kind hospitality 
at the Institute for Theoretical Physics of Tuebingen University. I 
would like to acknowledge also valuable help of G. Hoehler, his 
criticism of the early version of the manuscript, and useful 
discussions during my visit to the Institute for Theoretical 
Particle Physics of Karlsruhe University. This work is supported 
in part by the Alexander von Humboldt-Stiftung with a 
Forschungsstipendium and Neveu-INTAS and INTAS with Grants 
Nos. 93-0023 and 93-0079.

\newpage 

\begin{center}
{\bf FIGURE\ CAPTIONS}
\end{center}

{\bf Fig. 1}. Pictorial representation of the unitarity relations for
isovector nucleon form factors in the two-pion approximation for the
intermediate states between the photon and the nucleon-antinucleon pair. The
imaginary parts of the nucleon form factors are expressible in terms of a
product of the pion form factor and the $t$-channel $\pi N$-scattering
amplitudes.

{\bf Fig. 2}. (a,b) The values $F_{iv}(0)-F_{iv}(-t_0)$ (left scale) 
versus the number $N$ of the
weight coefficients $c_{ij}$ fixed by a requirement of minimum of the total
errors $\epsilon_{tot}$ originating from the unknown high-energy 
part of the integrals
and finite precision of the real parts of the $t$-channel $p$-wave
scattering amplitudes. (c,d) The ratios $Im\{{(F_{iv}(m_\rho ^2)\ 
-\ F_{iv}(-t_0))/F_\pi ({%
m_\rho ^2})\}}$ (right scale) at the $\rho $-meson peak versus 
the number $N$ of the
weight coefficients $d_{ij}$ fixed by a requirement of minimum of the total
errors $\delta_{tot}$ originating from the unknown high-energy part of the
integral in Eq.(3.9) and finite precision of the real part of the $t$%
-channel $p$-wave scattering amplitudes. The results of Ref. \cite{HP} 
(dashed regions) are plotted for comparison.

{\bf Fig. 3}. (a,b). Real part of the ratios between the nucleon form 
factors and the
pion form factor, $Re\{{(F_{iv}(t)\ -\ F_{iv}(-t_0))/F_\pi (t)\}}$, 
versus $t$
with errors $\epsilon_{tot}$ for $N=6$ (left scale). (a,b) Imaginary part 
of the 
ratios between the nucleon form factors
and the pion form factor, 
$Im\{{(F_{iv}(t)\ -\ Fiv(-t_0))/F_\pi (t)\}}/\frac{%
k^3}{\sqrt{t}}$, versus $t$ with errors $\delta_{tot}$ for $N=6$ 
(right scale). The results of Ref. \cite{HP} 
(dashed regions) are plotted for comparison.

\newpage

TABLE 1. Comparison of predictions of different methods at two points 
$t=0$ and $t=m_\rho ^2$. In Ref. \cite{LLME}, the estimates are obtained 
in the first order of the differential approximation. The errors 
(presently unknown) can be estimated 
by computing the next order of the differential approximation. The errors of 
Ref. \cite{HP} are evaluated from the claimed accuracy $\pm15\%$ 
for the amplitudes $f_{\pm}^1(m_{\rho }^2)$. Two sets of values are given on 
the basis of our calculations: shown in the line (I) are values 
corresponding to the maximum number of the weight coefficients $N = 12$. 
In the 
line (II) predictions obtained by statistical averaging the data from $N = 3$ 
to $12$ are given. In the line (I), the errors are theoretical ones, 
$\epsilon _{tot}$ and $(k^3/\sqrt{t})\delta _{tot}$ (see the text), 
while in the lower 
line the errors are statistical ones. The reduction of the errors 
for the case of the form factor $F_{2v}(t)$ occurs, since the variation of 
the results for different $N$ in this case is significantly smaller than 
the theoretical 
errors, $\epsilon _{tot}$ and $(k^3/\sqrt{t})\delta _{tot}$. 

$
\begin{array}{ccc}
\begin{array}{c}
Ref. \\ 
 \\ 
\cite{LLME} \\ 
\cite{HP} \\ 
This \; work \; (I) \\ 
This \; work \; (II) 
\end{array}
& {\ 
\begin{array}{c}
{(F_{iv}(0)\ -\ F_{iv}(-t_0))/F_\pi (0)} \\ 
\begin{array}{cc}
i = 1 & i = 2 \\ 
1.15 & 3.09 \\ 
 0.52\pm 0.08 & 3.15\pm 0.57 \\ 
-0.36\pm 0.35 & 2.73\pm 0.50 \\ 
-0.03\pm 0.37 & 2.78\pm 0.12 
\end{array}
\end{array}
} & {\ 
\begin{array}{c}
Im\{ 
{(F_{iv}(m_\rho ^2)\ -\ F_{iv}(-t_0))/F_\pi (m_\rho ^2)\}} \\ {\ 
\begin{array}{cc}
i = 1 & i = 2 \\ 
0.05 & 0.09 \\ 
0.20\pm 0.09 & 1.00\pm 0.21 \\ 
0.39\pm 0.07 & 1.08\pm 0.07 \\ 
0.34\pm 0.05 & 1.07\pm 0.01 
\end{array}
} 
\end{array}
} 
\end{array}
$

\end{document}